\documentclass[aps,prl,twocolumn,showpacs,superscriptaddress]{revtex4}

\usepackage{graphicx}
\usepackage{subfigure}
\usepackage{amsmath} 

\begin{document}

\title{Local current injection into mesoscopic superconductors \\for the manipulation of quantum states}

\author{M. V. Milo\v{s}evi\'{c}}
\affiliation{Departement Fysica, Universiteit Antwerpen,
Groenenborgerlaan 171, B-2020 Antwerpen, Belgium}

\author{A. Kanda}
\affiliation{Institute of Physics and TIMS, University of Tsukuba,
Tennodai, Tsukuba, Ibaraki 305-8571, Japan}

\author{S. Hatsumi}
\affiliation{Institute of Physics and TIMS, University of Tsukuba,
Tennodai, Tsukuba, Ibaraki 305-8571, Japan}

\author{F. M. Peeters}
\affiliation{Departement Fysica, Universiteit Antwerpen,
Groenenborgerlaan 171, B-2020 Antwerpen, Belgium}

\author{Y. Ootuka}
\affiliation{Institute of Physics and TIMS, University of Tsukuba,
Tennodai, Tsukuba, Ibaraki 305-8571, Japan}

\date{\today}

\begin{abstract}
We perform strategic current injection in a small mesoscopic
superconductor and control the (non)equilibrium quantum states in an
applied homogeneous magnetic field. In doing so, we realize a
current-driven {\it splitting of multi-quanta vortices},
current-induced transitions between states with {\it different
angular momenta}, and current-controlled switching between otherwise
{\it degenerate quantum states}. These fundamental phenomena form
the basis for discussed electronic and logic applications, and are
confirmed in both theoretical simulations and
multiple-small-tunnel-junction transport measurements.
\end{abstract}

\pacs{73.23.-b, 74.78.Na, 85.25.Hv}

\maketitle

Preparation and manipulation of discrete quantum states are crucial
for applications of quantum physics in nanoscale electronics,
particularly switching devices and memories. The needed quantum
states with discrete conductance levels, and the external control of
those, have been recently realized in solid electrolytes
\cite{terabe}, graphene \cite{li} (both voltage biased),
light-driven molecular switches \cite{delvalle}, and magnetic
nanowires tuned by magnetic field \cite{dugaev}. With a similar
goal, the current-driven processes started to attract immense
attention since the demonstration that the magnetization state of a
nanomagnet can be influenced directly by electrical current
\cite{krause}. The key convenience is that locally applied current
enables switching of {\it individual} submicron elements even in
integrated electronic circuits.

The latest example of the fundamental role of current injection in
low-dimensional physics is the achieved electric flipping of the
magnetic vortex core \cite{yamada}, without change in the vortex
chirality. Vortices, rotational flow of currents or matter with a
characteristic cavity at the center, can be found in many subfields
of quantum physics. In small magnetic elements, vortices owe their
existence to shape anisotropy and demagnetizing fields
\cite{cowburn}. In semiconducting quantum dots the rotation of
electrons is induced by external magnetic field and vortices may
form if this rotation is strong \cite{saarikoski}. Spontaneous
vortex nucleation has been recently observed even in
exciton-polariton condensates \cite{wouters}, a composite boson
system resembling superfluidity in Bose-Einstein condensates in
which vortices are readily found \cite{BEC}. Still, vortex matter is
most intrinsic to superconductors in a magnetic field. It is
particularly rich in samples comparable to the coherence length
$\xi$ and/or penetration depth $\lambda$ \cite{geim}, where vortex
states become strongly influenced by sample geometry \cite{baelus}.
Due to strong lateral confinement, vortices may even merge into a
multi-quanta (``giant'') vortex, otherwise unstable in an open
geometry \cite{schw,kanda}. Such vortices contain multiple phase
change of $2\pi$ encircling the central cavity, and are also found
in rotated superfluids and in strongly confined quantum dots in a
magnetic field \cite{saarikoski1}.
\begin{figure}[b]
\includegraphics[width=0.85\linewidth]{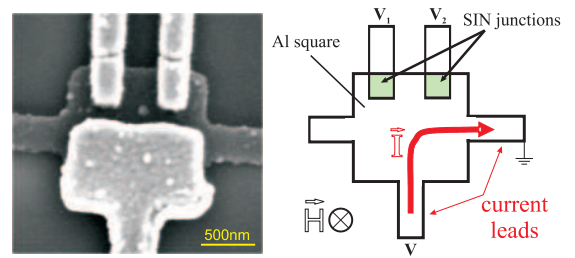}
\caption{SEM image of the sample - the mesoscopic superconducting
square in a transport structure, with two tunnel junctions on top,
as illustrated on the right (indicated directions of field and
current are denoted positive). The details of fabrication can
be found in Ref. \cite{kanda}.\label{Fig1}}
\end{figure}

Vortices not only represent a key feature of topologically confined
quantum-mechanical systems but may also be of use for electronic
device applications. It has been shown recently that each change in
the core of magnetic vortices gives rise to a measurable electric
signal \cite{prosandeev}. Changes in vortex configurations in
quantum dots intuitively lead to distinct features in electronic
transport \cite{gramila}. In superconductors, the quasiparticle
excitations inside vortices \cite{cdg} form coherent
quantum-mechanical states that depend on the number and arrangement
of trapped flux quanta. As a consequence, the sample conductance
measured along the vortex lines is determined by the transparency of
the given vortex configuration \cite{melvin}. Therefore, with
changing magnetic field and thanks to the consequent change in
vortex states, different levels of conductance can be achieved and a
quantum switch is realized. Up to date, varied magnetic field
remains the only tool for control of the angular momentum of vortex
states and their configuration. In this Letter, we present a new
method for vortex manipulation in small superconductors - by
strategically applied current. We demonstrate the first
electronically triggered splitting of the giant vortex as a
fundamental novelty, controlled transitions between vortex states of
different angular momentum for a given magnetic field (i.e. a
current-driven quantum switch), as well as the manipulation of
degenerate vortex equilibria from one to another, useful for logic
applications.

Our sample is a $d=40$ nm thick, $1.1\times1.1$ $\mu$m$^2$ Al
superconducting square designed for the
multiple-small-tunnel-junction (MSTJ) measurement \cite{kanda}, in
which two normal-metal (Cu) leads (width 0.25 $\mu$m) are connected
to the top side of the sample through tunnel junctions (see Fig.
\ref{Fig1}). In addition, the square has three direct Al leads
(width $0.3$ $\mu$m) centered at remaining sample sides, two of
which are used for current ($I$) injection. The coherence length was
estimated to $\xi(0)=150-190$ nm, and the superconducting transition
temperature $T_c$ was 1.35 K. The magnetic field sweep rate was 3
mT/min. In the MSTJ measurement, a small constant current (typically
1 nA) is applied to both junctions and the junction voltages, $V_1$
and $V_2$, are measured simultaneously as a function of the magnetic
field or the injected current $I$. An observed voltage reflects the
local density of states and also the local supercurrent density
underneath the junction, because the energy gap decreases with
increasing supercurrent density. As a result, voltage jump
corresponds to a transition between different vortex states.
Furthermore, the comparison of voltage signals at two junctions
indicates the symmetry of the circulating current, and thereby
$V_1\neq V_2$ points out the multi-vortex configuration, while
$V_1=V_2$ suggests a giant-vortex state.

To characterize the stationary and dynamic properties of the device,
we solved numerically the time-dependent Ginzburg-Landau (TDGL)
equation \cite{Kramer}
\begin{equation}
\small \frac{u}{\sqrt{1+\Gamma^2|\psi|^2}} (\frac {\partial
}{\partial t} + i\varphi +
\frac{\Gamma^2}{2}\frac{\partial|\psi|^2}{\partial t})\psi =(\nabla
-i\mathbf{A})^2 \psi +(1-|\psi|^2)\psi \label{tdgl1}
\end{equation}
coupled with the equation for the electrostatic potential $\Delta
\varphi = {\rm div}\left(\Im(\psi^*(\nabla-{\rm i}{\bf
A})\psi)\right)$. Here, the distance is measured in units of the
coherence length $\xi$, $\psi$ is scaled by its value in the absence
of magnetic field $\psi_{0}$, time by $\tau_{GL}=2T\hbar
\big/\pi\psi_0^2$, vector potential ${\bf A}$ by $c\hbar\big/
2e\xi$, and the electrostatic potential by
$\varphi_0=\hbar\big/2e\tau_{GL}$. $\Gamma=2\tau_E \psi_{0}/\hbar$,
with $\tau_E$ being the inelastic electron-collision time (for Al
samples, $\tau_E\sim 10$ ns gives $\Gamma\approx10^3$). Parameter
$u=5.79$ is taken from Ref. \cite{Kramer}. Note that in Eq.
(\ref{tdgl1}) the screening of the magnetic field is neglected, as
our sample is sufficiently thin to exhibit effective type-II
behavior (therefore ${\bf A}=\frac{1}{2}B(y,-x,0)$). The current
leads (see Fig. \ref{Fig1}) were simulated by imposing $-\nabla
\varphi=j_i$, where $j_i$ is the injected current density in units
of $j_0=c\Phi_0\big/8\pi^2\Lambda^2\xi$, $\Lambda=\lambda^2/d$. At
the remainder of the sample edges, Neumann boundary condition was
used ($j_{\bot}=0$).
\begin{figure}[t]
\includegraphics[width=\linewidth]{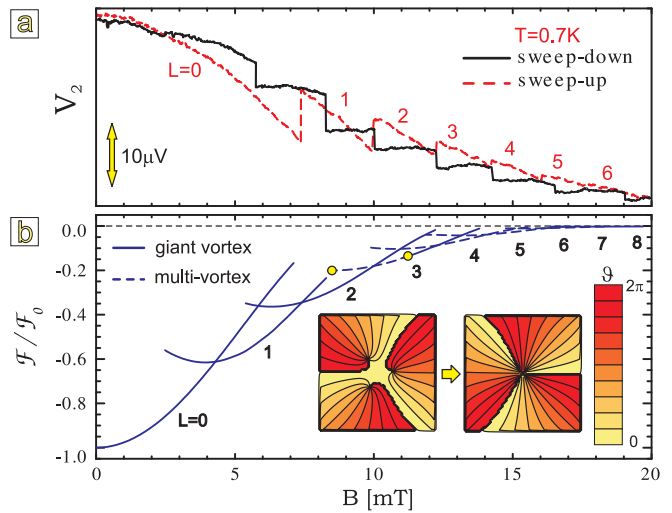}
\caption{Measured voltage (a) and calculated free energy (b) at
$T=0.7$ K. Insets show phase contourplots of the superconducting
order parameter for $L=3$ states (multi- and giant vortex) marked by
open dots in free energy curves.\label{Fig2}}
\end{figure}

Fig. \ref{Fig2} shows the measured voltage and calculated Helmholtz
free energy of the sample as a function of applied magnetic field
($\mathcal{F}=\frac{1}{V}\int_V |\psi|^4dV$, in units of
$\mathcal{F}_0=H_c^2/8\pi$, where $V$ is the sample volume). In Fig.
\ref{Fig2}(b) different energy levels correspond to states with
different angular momentum, i.e. vorticity $L$. Here taken
$\xi(0)=152$ nm and $T_c=1.34$ K provided the best agreement with
experimental vortex penetration/expulsion fields [see Fig.
\ref{Fig2}]. As already pointed out, some vortex states may exhibit
ring-like configuration of individual vortices \cite{schw}, i.e. a
multi-vortex [dashed lines in Fig. \ref{Fig2}(b)], while in others
all flux quanta coalesce into a single giant-vortex (solid lines).
At higher temperatures, the characteristic lengthscales grow and
effective confinement favors giant-vortex states.

\begin{figure}[t]
\includegraphics[width=\linewidth]{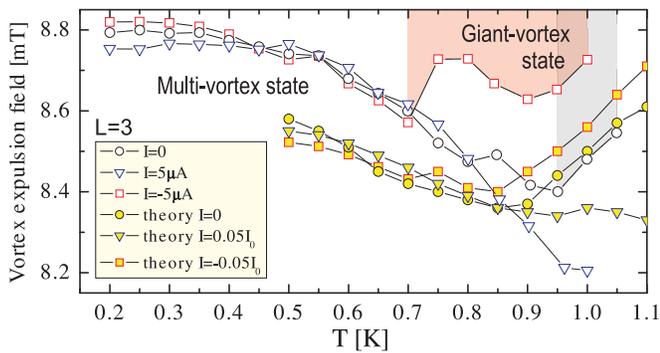}
\caption{Experimentally and theoretically obtained vortex expulsion
field as a function of temperature for the $L=3$ vortex state
without and with applied current. Different behavior of the curves
indicates multi- or giant- (shaded areas in experimental curves)
vortex state. \label{Fig3}}
\end{figure}

The distinction between the giant- and multi-vortex states in
mesoscopic superconductors can also be made by monitoring of the
vortex expulsion field $B_e$ for a given state as a function of
temperature, as detailed in Ref. \cite{Ben} (decreasing $B_e(T)$
tendency is found for a multi-vortex vs. increasing $B_e(T)$
behavior for a giant-vortex state). Our measurements and
calculations indicated that for e.g. $L=3$ vortex state, the
giant-vortex form is favorable above 0.9 K. In what follows, we
investigate the influence of applied current on the shape of the
vortex state. Following the scheme shown in Fig. \ref{Fig1}, we
applied weak dc current in the direction down-to-right, such that
vortices experience Lorentzian-type of force (${\bf F}={\bf
j}_i\times{\bf \Phi}$) approximately towards top-left corner of the
sample. Due to the spatial inhomogeneity of the force and the
interaction with the sample boundaries, it can be energetically
favorable for the giant-vortex to separate into individual ones and
accommodate the imposed conditions. We achieved this {\it
giant-to-multi vortex splitting} in both theory and experiment at
temperatures above $0.95$ K and this is the first such result in
vortex physics. As shown in Fig. \ref{Fig3}, for e.g. $T=1.0$ K, we
found a giant-vortex in the absence of applied current, but
multi-vortex when even a weak current of $5~\mu$A was applied. In
supplementary material \cite{supplmat}, we also show the simulated
dynamics of the giant-vortex splitting under applied drive at $T=1$
K. Note however that in the case of opposite current ($I=-5~\mu$A),
the drive `drags' vortices towards the bottom-right corner, where
superconductivity is strong. This compresses the previously existing
multivortex, and favors the giant-vortex state (see corresponding
experimental curve in Fig. \ref{Fig3} for $0.7<T<0.95$ K). In
addition, the vortex expulsion field is higher compared to $I=0$
case, since the specific Lorentzian drive restricts access of
vortices to the sides of the sample.

\begin{figure}[t]
\includegraphics[width=\linewidth]{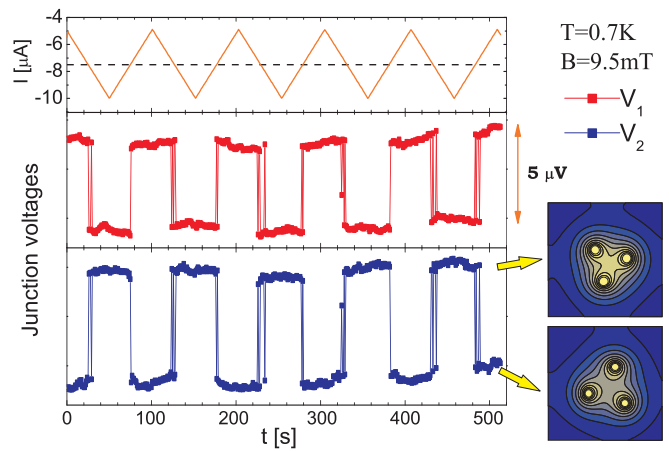}
\caption{Switching between degenerate vortex states (see insets) by
alternating current ($|I|<>-7.5~\mu A$), demonstrated experimentally
through measured voltages at junctions 1 and 2 (see Fig.
\ref{Fig1}).\label{Fig4}}
\end{figure}

The above splitting of a giant-vortex by injected current is a
non-equilibrium process, as giant- and multi-vortex realization of
the same $L$-state cannot coexist. However, for a given multi-vortex
state, there can be maximum $n$ energetically degenerate vortex
configurations with respect to $n$ sample boundaries. This fact
forms a base for the fluxonic cellular automata \cite{fca}, where
e.g. two vortices in a square can occupy two alternative diagonal
positions, forming states that can be labeled as logic `0' and `1'.
One of the essential problems in cellular automata is the successful
preparation of inputs, i.e. the feasible manipulation of bits, and
this can be realized by local current injection. In Fig. \ref{Fig4},
we show the manipulation of the $L=3$ multi-vortex state at $T=0.7$
K. For sufficiently strong negative current, two vortices are
attracted to the bottom-right corner rather than just one, which
rotates the vortex structure by c.a. $30^{\circ}$. In other words,
the applied current enables the transition between otherwise
degenerate states, and the switching is fully reversible. This is
demonstrated in Fig. \ref{Fig4}, where we realized multiple
switching back and forth between two degenerate states by an
alternating current. The rotation of the vortex configuration is
verified by the alternating voltages measured at the two tunnel
junctions.

\begin{figure}[tb]
\includegraphics[width=\linewidth]{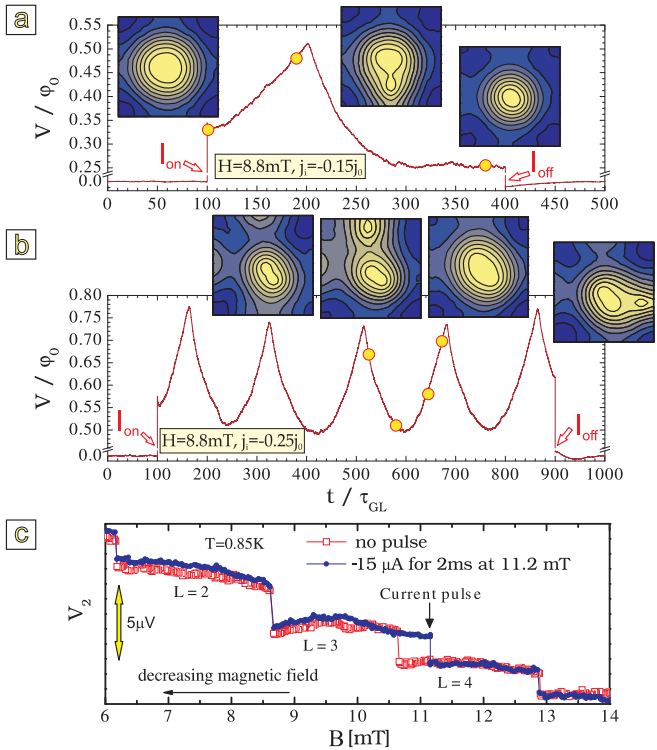}
\caption{Calculated voltage between current leads vs. time (at
$T=1.0$ K), for the case of (a) pullout of a flux-quantum
(transition from $L=3$ to $L=2$), and (b) active vortex penetration
and expulsion, with average vorticity $L=3$ (insets depict snapshots
of the Cooper-pair density, see corresponding animations in
\cite{supplmat}). (c) Experimentally measured voltage vs. decreasing
applied field, prior and after the current-induced $L=4\rightarrow
3$ transition.\label{Fig5}}
\end{figure}

Obviously, here reported phenomena are sensitive to the sample
parameters, but crucially depend on the magnitude of the applied
current. The larger current exerts larger force on vortices, which
may expel some of them and/or help nucleate new ones in cases when,
roughly speaking, local current on one of the sample edges exceeds
the depairing current. The first scenario is particularly
interesting at applied magnetic field which stabilizes several
vortex states (see e.g. Fig. \ref{Fig2}, for $B=12$ mT). In such a
case, one can induce {\it controlled transitions between states of
different vorticity} by applying current of needed magnitude! We
show this in Fig. \ref{Fig5}(a), where one vortex is pulled {\it
out} of the ground-state $L=3$ giant-vortex (by $I<0$), and double
vortex ($L=2$) remains stable after the current is switched off.
Successful switching manifests as a peak of voltage $V$ vs. time. If
the same state was exposed to even larger current, vortices are
sequentially expelled from the $L=3$ state, but also new vortices
appear at the opposite edge of the sample. Eventually, a `dynamic
equilibrium' is established (as shown in Fig. \ref{Fig5}(b) and in
supplementary animations \cite{supplmat}), exhibiting clear
periodicity in the $V(t)$ signal linked to entry and exit of
vortices. Latter oscillations of voltage are tunable by applied
current, but in the present (Al) sample they are mostly in the GHz
frequency range ($\tau_{GL}\approx 0.2\big/(1-T/T_c)$ ps).

In Fig. \ref{Fig5}(c) we show the experimental evidence for
current-driven transitions between states of different angular
momentum, at $T=0.85$ K \footnote{At $T>0.9$ K, low measured voltage
and low critical current of the superconducting leads hamper the
precision of the measurement.}. After decreasing magnetic field to
$B=11.2$ mT, the $L=4$ to $L=3$ transition was induced by applied
current of $I=-15$ $\mu$A, analogously to the scenario shown in Fig.
\ref{Fig5}(a). Voltage measured in further decreasing field shows
that $L=3$ state indeed remains stable. Therefore, our sample can
act as an excellent quantum switch, where applied current
opens/closes the ballistic channel for quasiparticles through the
vortex core (see Ref. \cite{melvin}).

\begin{figure}[tb]
\includegraphics[width=0.9\linewidth]{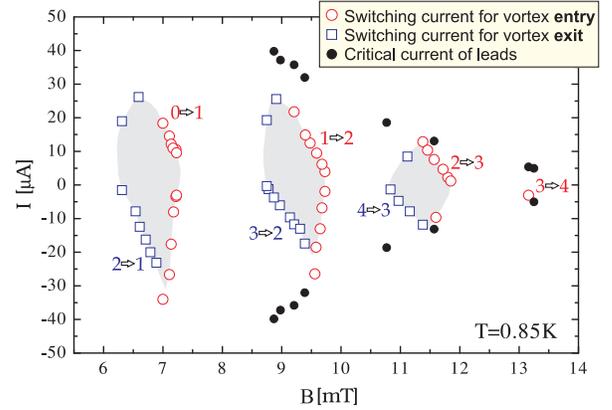}
\caption{Minimal switching current for states with $0<L\leq 4$
(experiment done at 0.85 K), with distinction between the
$L\rightarrow L-1$ and $L\rightarrow L+1$ processes (maximal current
in experiment is limited by the critical current of the leads). The
quantum switch was not operational in the shaded areas.\label{Fig6}}
\end{figure}

Latter $L=n\leftrightarrow n+1$ transitions can be achieved for both
polarities of the applied current, as one fluxon can be either
`pushed' or `pulled' out of the sample. In Fig. \ref{Fig6} we give
the experimentally obtained values for the threshold current needed
for enforced transitions between states with vorticity $0<L\leq 4$
at $T=0.85$ K, both for increasing and decreasing magnetic field.
The typical values of minimal current needed for the switching were
in the $|I|<30~\mu$A range for $T=0.85$ K, and grew significantly
with decreasing temperature (e.g. $|I|<80~\mu$A range for $T=0.7$
K). Note that in some cases it becomes irrelevant if we perform
vortex penetration or expulsion - e.g. at $B=7$ mT, applied current
of $I>30~\mu$A will provide $L=1$ state, regardless of the initial
vortex state ($L=0$ or $L=2$).

To summarize, we realized electronic control of the quantum states
in a mesoscopic superconductor by strategic injection of dc current.
Out of equilibrium, we performed the splitting of a giant,
multi-quanta vortex into individual vortices. In equilibrium, we
demonstrated the current-induced transitions between states with
different angular momenta, which also has a step-like effect on
local ballistic conductance and is applicable as a quantum switch.
Finally, switching between dual equilibria was also achieved, most
useful for the fluxonic cellular automata. Local current injection
therefore proves to be a powerful method for manipulation and
preparation of quantum states on a submicron scale, and further work
is needed to test the analogies in other quantum systems.

This work was supported by the Flemish Science Foundation (FWO-Vl),
the Belgian Science Policy (IAP), JSPS/ESF-NES, and the Sumitomo
and the Hitachi-Kurata
Foundations.

\end{document}